# Low cognitive reflection predicts honesty for men but not for women

Valerio Capraro and Niko Peltola

Middlesex University London

Contact author: V.Capraro@mdx.ac.uk

**Abstract**

Previous experiments have explored the effects of gender and cognitive reflection on dishonesty separately. To the best of our knowledge, no studies have investigated potential interactions between these two factors. Exploring this interaction is important because previous work found that males tend to be both more deliberative than females. Therefore, it is possible that the gender effect on dishonesty is moderated by cognitive reflection. Here we report a large online experiment (N = 766) where subjects first have a chance to lie for their benefit and then take a Cognitive Reflection Test (CRT). We find a significant interaction between gender and CRT score such that lack of deliberation promotes honesty for men but not for women. Additional analyses highlight that this effect is mainly driven by men whose answers in the CRT are neither intuitive nor deliberative, who happen to be particularly honest in our deception game.

# Introduction

Dishonesty has a negative impact on people, companies, and the society as a whole. For example, according to a 2001 study by the Internal Revenue Service (IRS), the "tax gap" between tax owed and tax paid, which results mostly from tax evasion, is between $312 billion and $353 billion annually, which corresponds to a noncompliance rate between 15% and 16.6% (Herman 2005).

Understanding which factors influence dishonesty has therefore inspired a great deal of work (Gneezy, 2005; Mazar, Amir & Ariely, 2008; Fischbacher & Föllmi-Heusi, 2014). Among the most studied factors, there are gender and cognitive reflection.

### *The effect of gender on dishonesty*

Dreber & Johannesson (2008) found that men lie more than women in the context of *self-serving lies* (also called *black lies*). This result was replicated by Friesen and Gangadharan (2012), while, instead, Childs (2012) found no gender differences. Erat & Gneezy (2012) observed that the consequences of lying matter: women are more likely than men to tell *altruistic white lies* (lies that benefit another person at a cost for the liar), but men are more likely than women to tell *Pareto white lies* (lies that benefit all parties involved). However, the latter result was not replicated by Cappelen, Sørensen & Tungodded (2013). Biziou-van-Pol, Haenen, Novaro, Occhipinti-Liberman and Capraro (2015) also found no gender differences in the domain of Pareto white lies; additionally, in contrast to Erat & Gneezy (2012), they found that men are more likely than women to tell altruistic white lies. Given this mixed evidence, researchers have recently turned to meta-analytic techniques: Abeler, Nosenzo and Raymond (in press) analyzed the decisions of over 32,000 subjects and found that men lie more than women, but they did not control for the consequences of lying; Capraro (2018) analyzed the decisions over 8,000 subjects and found that men tell more black lies and more altruistic white lies than women, while results are inconclusive in the case of Pareto white lies.

### *The effect of cognitive reflection on dishonesty*

A handful of papers explored the effect of cognitive reflection, measured through the Cognitive Reflection Test (CRT), on dishonesty, finding mixed results. Gino and Ariely (2012) found CRT not to be correlated to honesty. However, Fosgaard et al (2013) found CRT to be positively associated with cheating. Interestingly, Ruffle and Toble (2016) found the opposite, that high CRT predicts honesty. This mixed literature is mirrored also in studies manipulating cognitive reflection mode. Some studies found that time delay (as a proxy for reflective thinking, Rand, Greene & Nowak, 2012; Capraro & Cococcioni, 2015; Merkel & Lohse, 2018) promotes dishonesty (Capraro, 2017; Capraro, Schulz & Rand, 2018; Lohse, Simon & Konrad, 2018), while others suggested it promotes honesty (Gunia et al, 2012; Shalvi et al, 2012), and yet another one reported no effect (Barcelo & Capraro, 2017). Also studies using conceptual priming of intuition, cognitive load, or ego-depletion found mixed results (Cappelen, Sørensen & Tungodded, 2013; Vershuere et al, 2018; Gino, Schweitzer, Mead & Ariely, 2011; van't Veer, Stel & van Beest, 2014).

### *Theoretical motivation for studying how gender and cognitive reflection interact in the context of honest behavior*

As mentioned earlier, previous works have investigated the effect of gender and cognitive reflection on honesty, but they did so separately. This is an important limitation because a line of recent research suggests that men tend to score higher than women in the CRT (Brañas-Garza, Kujal & Lenkei, 2015; Campitelli & Gerrans, 2014; Cueva et al, 2016; Pennycook, Cheyne, Koehler & Fugelsang, 2016; Sinayev & Peters, 2015; Ring, Neyse, David-Barett & Schmidt, 2016; Albaity, Rahman & Shahidul, 2014; Toplak et al, 2014; Primi, Morsanyi, Chiesi, Donati &

Hamilton, 2016). This raises an important question: Is the effect of cognitive reflection on honesty moderated by gender?

To the best of our knowledge, no studies have explored this question in the domain of honest behavior. Somewhat related, previous research has investigated the interaction between gender and experimental manipulation of cognitive processing in behavioral domains such as altruism and cooperation: Rand et al (2016) found, in a meta-analysis of 22 dictator game experiments, that promoting intuition versus deliberation favors altruism for women but not for men; whereas Rand (2017) found, in a meta-analysis of 67 studies, that promoting intuition versus deliberation favors cooperation for men and women alike.

Given this gap in the literature, here we aim at doing a first step in the direction of exploring the interaction between gender and cognitive reflection in the domain of honest behavior.

## Method

### *Measure of cognitive reflection*

To measure participants' cognitive reflection trait, we use the Cognitive Reflection Test (CRT). First developed by Frederick (2005), the CRT includes three questions characterized by the property that an automatic, intuitive answer typically pops up to people's mind. However, this answer is wrong and, in order to find the right answer, people have to overcome this automatic reaction. For example: A bat and a ball cost $1.10 in total; if the bat costs $1.00 more than the ball, how much does the ball cost? The intuitive answer is $0.10. A moment of reflection, however, rejects this answer: if the ball costs $0.10 and the bat costs $1.00 more than the ball, then the bat costs $1.10; thus, the bat and the ball together cost $1.20, and not $1.10 as assumed. The actual cost of the ball is indeed $0.05. The CRT score (number of correct answers in the CRT) has been shown to correlate with analytic cognitive style, measured using the Need for Cognition scale (Pennycook,

Cheyne, Koehler & Fugelsang, 2016). Thus, we take this number, the CRT score, as a measure of cognitive reflection.

*Measure of honesty*

To measure participants' honesty, we use a variant of Gneezy's (2005) Deception game that has been formulated by Biziou-van-Pol et al (2015). In this variant, subjects are each matched with an anonymous *receiver* and are informed that they have been randomly assigned to either Group 1 or Group 2, and that they will be asked to declare which group they have been assigned to. They can choose between: "I have been assigned to Group 1" or "I have been assigned to Group 2". If they report the true number, they will get $0.4 while the receiver gets $0.5; instead, if they report the other number, they will receive $0.5, while the receiver gets $0.4. Subjects are informed that the receiver will not be informed about their true group number, and about the payoffs corresponding to the available choices. Two comprehension questions, one regarding the choice maximizing their payoff and one regarding the choice maximizing the receiver's payoff, are asked before making a decision. Subjects failing either or both comprehension questions are automatically excluded from the survey. Thus, subjects play the game having a full understanding of its rules.

We chose this variant of the deception game to avoid having subjects that use *sophisticated deception*: telling the truth because they believe that the receiver will not believe them (Sutter, 2009). In our case, the receiver does not make any choice, and thus beliefs about the beliefs of the receiver do not play any role.

We chose this particular payoff structure, instead of more classical ones in which both players get the same payoffs when the sender tells the truth (e.g., Biziou-van-Pol et al, 2015; Cappelen et al, 2013; Erat & Gneezy, 2012; Gneezy, 2005), because previous research shows that low CRT is associated with more egalitarian choices (Capraro et al, 2017). This could have been a source of confound in case honesty had brought the same payoff to both players, such that we would have not been able to conclude that any effect of CRT on honesty would be driven by lying aversion: it could be driven by the payoff consequences of the available actions. Additionally, Capraro et al (2017) also show that high CRT is associated with socially efficient choices. This is why we opted for a payoff structure such that the sum of the payoffs of the two players and their absolute difference is constant across choices. Finally, Capraro et al (2017) also find that low CRT is associated with spitefulness, while high CRT is associated with payoff maximization. This does not create any confound in our setting, because the spiteful choice coincides with the payoff maximizing choice.

*Data collection and procedure*

We recruited subjects on Amazon Mechanical Turk (AMT) (Arechar, Gächter & Molleman, 2018; Brañas-Garza, Capraro & Rascón-Ramírez, 2018; Goodman, Cryder & Cheema, 2013; Paolacci, Chandler & Ipeirotis, 2010; Horton, Rand, Zeckhauser, 2011; Paolacci & Chandler, 2014). Subjects were living in the US at the time of the experiment and earned $0.50 for completing the survey, plus an additional bonus depending on the choice made in the Deception game.

Subjects first participated in the Deception game, then took the CRT, and finally completed a demographic questionnaire, at the end of which they received a completion code, through which they could submit the survey on AMT and claim for their payment. All subjects made a decision in the Deception game. After the survey was completed, we randomly created pairs of subject, and, each subject, was paid two bonuses, one as a chooser and one as a receiver (when making their choice, participants were not informed that they will receive a payoff also as a receiver to avoid having this impact their decision). Experimental instructions are reported in the appendix.

**Results**

*Participants and variables*

A total of 766 participants passed the comprehension questions and participated in the experiment. Table 1 reports descriptive statistics of all the variables in our study. *Female* is a dummy variable that takes value 1 if the participant reports that she is a female. *Age* is self-explanatory. *Education* takes value 1 if the participant reports that their higher highest level of education completed is "less than high school diploma"; Education = 2 corresponds to "high school diploma"; Education = 3 corresponds to "vocational training"; Education = 4 corresponds to "attended college"; Education = 5 corresponds to "bachelor's degree"; Education = 6 corresponds to "graduate degree"; Education = 7 corresponds to "unknown" (no participants have actually reported Education = 7). *Honesty* takes value 1 if the subject told the truth in our deception game. *CRT* counts the number of correct answers in the Cognitive Reflection Test. *CRT_intuitive* counts the number of intuitive answers in the Cognitive Reflection Test. *CRT_residual* counts the number of answers in the Cognitive Reflection Test that are neither intuitive nor deliberative.

|  | mean | median | std | min | max |
|---|---|---|---|---|---|
| female | 0.51 | 1 | 0.50 | 0 | 1 |
| age | 37.4 | 34 | 11.44 | 18 | 82 |
| education | 4.42 | 5 | 1.17 | 1 | 6 |
| honesty | 0.30 | 0 | 0.46 | 0 | 1 |
| CRT | 1.58 | 2 | 1.22 | 0 | 3 |
| CRT-intuitive | 2.15 | 3 | 1.13 | 0 | 3 |
| CRT-residual | 0.19 | 0 | 0.46 | 0 | 3 |

*Table 1. Descriptive statistics of the variables.*

### *Main effect of gender on honesty*

We start by looking at the main effect of gender on honesty. Average honesty among males was 0.31, while average honesty among females was 0.28. The difference is not statistically significant (logit regression predicting *Honesty* as a function of *Female*, p = 0.386). This result is robust for controlling on age and education (p=0.330). In sum, we find no gender differences in deception in our task.

### *Main effect of cognitive reflection on honesty*

Next, we look at the main effect of cognitive reflection on honesty. Logistic regression predicting *Honesty* as a function of *CRT* finds no statistically significant effect (p=0.187). When including control on female, age, and education, we find a marginally significant negative effect of *CRT* on *Honesty* (coeff = -0.121, z = -1.77, p = 0.076). Similar results hold if we replace *CRT* with *CRT-intuitive*, however, this time, the effect of *CRT-intuitive* on honesty is not statistically significant both when we do not include control on female, age, education (p=0.232) and when we include control (p=0.104). Thus, in our sample, we find very little evidence of a main effect of cognitive reflection on honesty.

### *Main effect of gender on cognitive reflection*

Now, we look at the main effect of gender on cognitive reflection. Linear regression predicting *CRT* as a function of *Female* finds a highly significant negative effect (coeff = -0.564, t = -6.57, p < .001). This result is robust after controlling for age and education (coeff = -0.589, t = -6.91, p < .001). Thus, in our sample, females tend to score lower than males in the cognitive reflection test.

### *The interaction between Cognitive Reflection and Gender*

Coming to our main research question, we now conduct logit regression predicting *Honesty* as a function of the following variables: *Female*, *CRT*, and their interaction *Female X CRT*. Results are reported in Table 2 (column (I) with no control on *Age* and *Education*, column (II) with control). We find that both female and *CRT* have a negative effect on *Honesty* (*Female:* coeff = -0.675, z = -2.58, p =0.010; *CRT:* coeff = -0.265, z = -2.76, p = 0.006). Interestingly, also the interaction *Female X CRT* is significant (coeff = 0.310, z = 2.33, p = 0.020). These results are robust after controlling for age and education (see Table 1, Column (II)).

|  | Honesty | |
|---|---|---|
|  | (I) | (II) |
| **Female** | -0.675** (0.262) | -0.714*** (0.264) |
| **CRT** | -0.265*** (0.096) | -0.284*** (0.112) |
| **Female X CRT** | 0.310** (0.133) | 0.314** (0.158) |
| **Age** |  | 0.009 (0.007) |
| **Education** |  | 0.049 (0.069) |
| **Constant** | -0.311 (0.202) | -0.827* (0.433) |
| **Observations** | 766 | 766 |

*Table 2. Logit regressions predicting honesty as a function of the highlighted variables. *: p < 0.10; **: p < 0.05; ***: p < 0.01.*

The significant interaction between *Female* and *CRT* suggests that *CRT* affects *Honesty* differently for males and females. To better understand this effect, we conduct logit regression predicting *Honesty* as a function of *CRT* for males and females separately. We find that *CRT* has a significant negative impact on *Honesty* among males (coeff = -0.265, z = -2.76, p = 0.006) but not females (coeff = 0.045, z = 0.48, p = 0.628). This result is clear also from Figure 1, where, to increase readability, we divide subjects into "high deliberation" (*CRT* = 2 or 3) and "low deliberation" (*CRT* = 0 or 1).

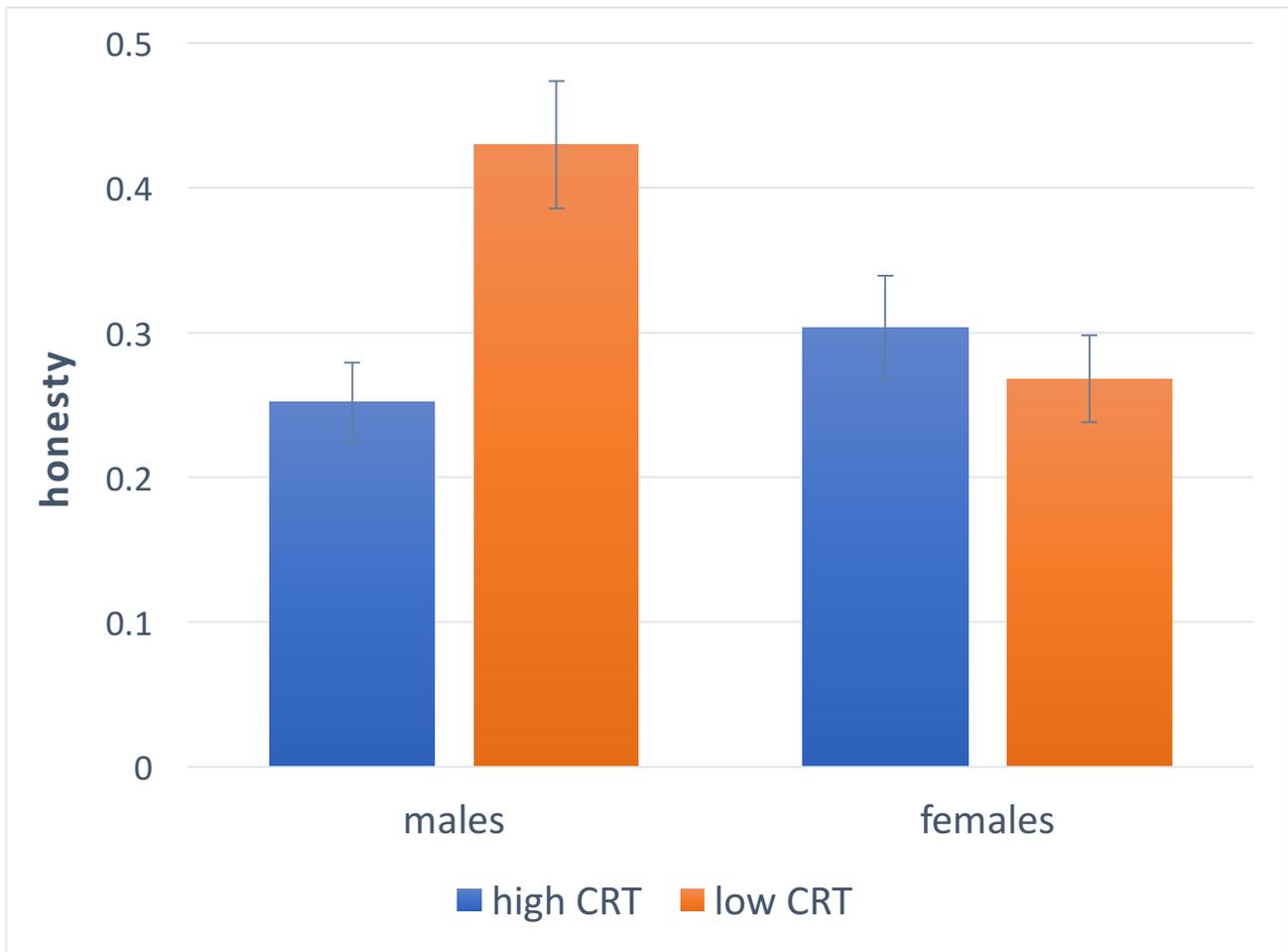

*Figure 1. CRT has a negative effect on Honesty among males but not females.*

In order to have a better understanding of this effect, we explore the impact of the co-variable *CRT-intuitive*, which counts the number of intuitive responses in the CRT, on honesty. We specify since now that this variable does not represent a reliable measure of intuitive cognitive style, at least when measured using the Faith in Intuition scale (Pennycook et al, 2016). With this in mind, this analysis will however allow us to have a better understanding of the aforementioned interaction between CRT and gender. We conduct logit regression predicting *Honesty* as a function of *Female, CRT-intuitive*, and the interaction *Female X CRT-intuitive*. Results are reported in Table 2. Interestingly, while we still find a significant effect of *CRT-intuitive* on honesty, this time the interaction *Female X CRT-intuitive* is not significant.

|  | Honesty | |
|---|---|---|
|  | (I) | (II) |
| **Female** | 0.051 (0.236) | 0.024 (0.237) |
| **CRT-intuitive** | 0.201** (0.100) | 0.219** (0.101) |
| **Female x CRT-intuitive** | -0.192 (0.137) | -0.195 (0.138) |
| **Age** |  | 0.009 (0.007) |
| **Education** |  | 0.049 (0.069) |
| **Constant** | -0.991*** (0.153) | -1.542*** (0.447) |
| **Observations** | 766 | 766 |

*Table 3. Logit regressions predicting honesty as a function of the highlighted variables. \*: p < 0.10; \*\*: p < 0.05; \*\*\*: p < 0.01.*

This analysis suggests that the aforementioned negative effect of *CRT* on honesty for men but not for women is driven by men who give answers in the CRT that are neither deliberative nor intuitive, who happen to be honest in the deception game. To strengthen this interpretation, we define a variable *CRT-residual* (we adopt the terminology from Cueva et al, 2016) which counts the answers in the CRT that are neither deliberative, nor intuitive, and we analyze whether this variable impacts the level of honesty while interacting with gender. Logit regression predicting *Honesty* as a function of *Female*, *CRT-residual*, and the interaction *Female X CRT-residual* shows a significant interaction (coeff = -0.917, z = -2.57, p = 0.010), such that providing neither intuitive nor deliberative answers in the CRT predicts honesty for males (coeff = 0.532, z = 2.22, p = 0.027) but not for females (coeff = -0.384, z = -1.46, p = 0.145). Moreover, logit regression predicting *Honesty* as a function of *Female*, *CRT*, and their interaction, gives a non-significant interaction effect when restricted to the set of subjects such that *CRT-residual=0* (p=0.124). Figure 2 represents these results, by reporting honesty as a function of whether subjects gave at least one answer in the CRT that was neither intuitive nor deliberative. (We opted for this representation of these results, because only 20 subjects have *CRT-residual = 2* or *CRT-residual = 3*).

Taken together, these findings demonstrate that lack of deliberation favors honesty for men but not for women, and that this effect is mainly driven by men who give answers that are neither intuitive nor deliberative.

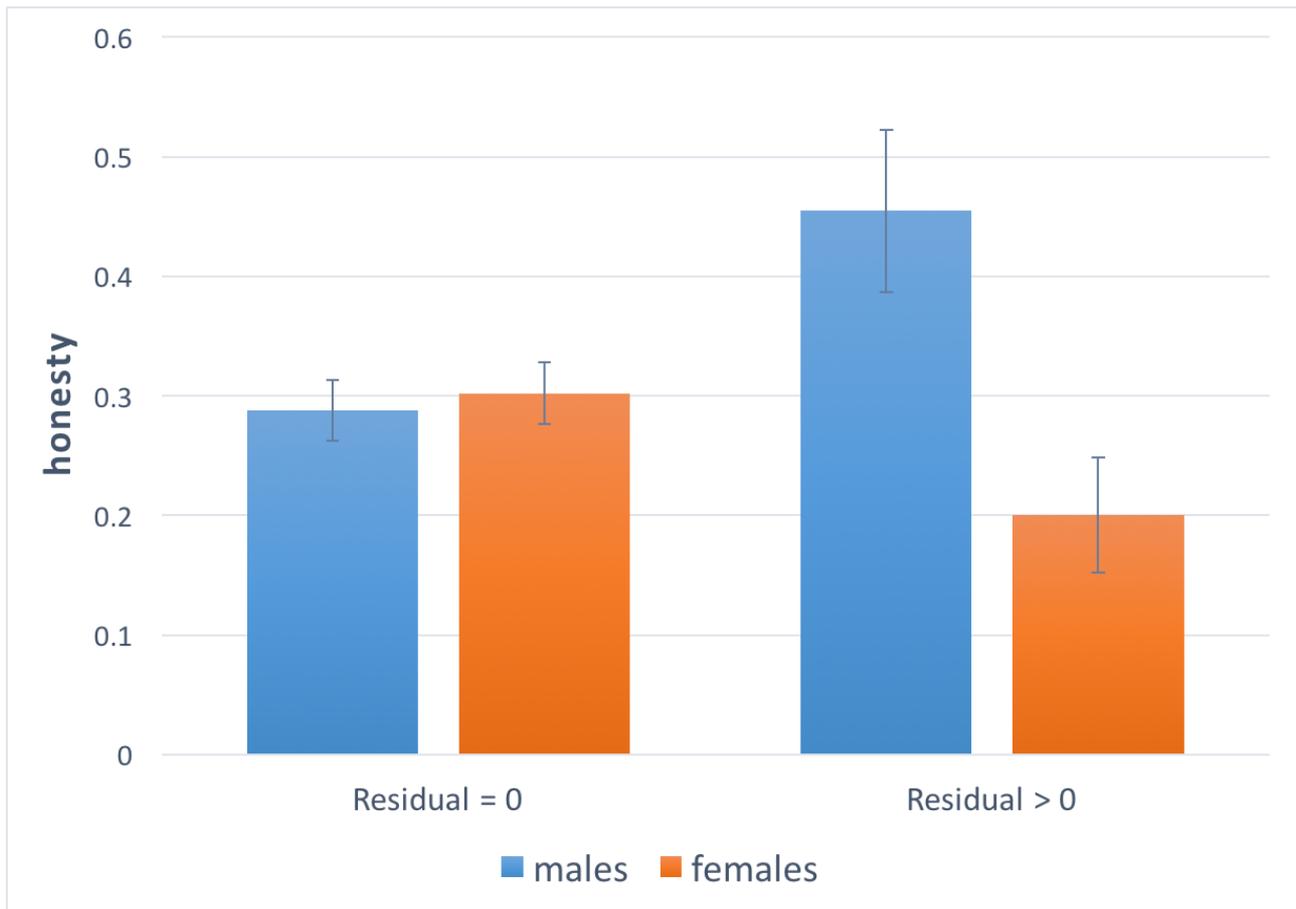

*Figure 2. Providing at least one answer in the CRT that is neither intuitive nor deliberative (Residual > 0) predicts honesty for males but not for females.*

**General discussion**

Understanding the effect of gender and cognitive reflection on honesty is a current topic of debate. However, to the best of our knowledge, no studies have explored the interaction between these two factors. Investigating this interaction is important because previous works have found that men score higher than women in the Cognitive Reflection Test, a fact that raises the question: Is the effect of cognitive reflection on honesty moderated by gender?

Given this gap in the literature, we have conducted a large (N=766) study exploring how gender and cognitive reflection interact in determining people's behavior in a deception game, using a large (N=766) sample of American subjects recruited online on Amazon Mechanical Turk. We found a significant interaction, such that lack of deliberation predicts honesty among men but not among women. Additional analyses showed that this effect is driven by men whose answers in the cognitive reflection test are neither intuitive nor deliberative.

Our results add to the growing body of literature on the effect of gender and cognitive reflection on honesty. Previous work found mixed results regarding the role of cognitive reflection on truth-telling: Fosgaard et al (2013) found a positive effect; Gino & Ariely (2012) found a null effect; Ruffle & Toble (2016) found a negative effect. We add to this line of research by showing that, at least in our context, low CRT is associated with honesty, but only among males. Why only among males? This is an important question, about which, at this stage of the research, we can only speculate. One thing to note, first of all, is that all our subjects have passed comprehension questions about the deception game: thus we can exclude the trivial explanation that males are more likely than females to get confused by the deception game (and by the CRT) – a hypothesis that, in principle, could have explained our results. Another explanation that we can exclude is that males are less deliberative than females and that lack of deliberation favors honesty in this task: in line with previous research and in contrast with this hypothesis, we find that females score significantly lower than males in our CRT. One possible explanation, instead, is that some males are more careless than females: although they understand the general rules of the game and pass the comprehension questions (which guarantee them the basic payment), they lose interest and, when they are asked to report the group they are assigned to, they tell the true group number not because they want to be honest, but just because it is easy; and, when they take the CRT, they are more likely to make "total mistakes" and give answers that are neither deliberative nor intuitive. We hope that further research might help clarify this point. In this light, also theoretical studies making predictions about when and why we should expect an interaction between gender and cognitive reflection will be of great help.

Regarding the main effects of gender and cognitive reflection on honesty, we found non-statistically significant results in both cases. Whether there is a main effect of gender on honesty has long been debated, from the early work by Dreber and Johannesson (2008), which found that males tell more self-serving lies than females. Since then, several studies have explored this question, by reporting mixed results (Biziou-van-Pol et al, 2015; Cappelen et al, 2013; Childs, 2012; Erat & Gneezy, 2012; Friesen & Gangaradhan, 2012). To shed light on this mixed evidence, researchers have recently turn to meta-analytic techniques: two independent meta-analyses have both found that males are more dishonest than males (Abeler et al, in press; Capraro, 2018), at least in the case of self-serving lies and altruistic white lies (i.e., lies that help another person to a cost for the liar). We add to this literature by showing that, at least in our context, there are no gender differences in deception. As for the main effect of cognitive reflection, to the best of our knowledge, only three empirical studies analyzed the correlation between cognitive reflection and honesty before us, and found mixed results: Fosgaard et al (2013) found a positive effect; Gino & Ariely (2012) found a null effect; Ruffle & Toble (2016) found a negative effect. We add to this line of research by showing that, at least in our context, there is no main effect of cognitive reflection on honesty.

Our results add also to the growing body of literature regarding gender differences in cognitive reflection. In line with previous literature (Brañas-Garza, Kujal & Lenkei, 2015; Campitelli & Gerrans, 2014; Cueva et al, 2016; Pennycook, Cheyne, Koehler & Fugelsang, 2016; Sinayev & Peters, 2015; Ring, Neyse, David-Barett & Schmidt, 2016; Albaity, Rahman &

Shahidul, 2014; Toplak et al, 2014; Primi, Morsanyi, Chiesi, Donati & Hamilton, 2016), also we found that females score less than males in the CRT.

Another important feature of our work is that we used a large (N=766) sample of subjects that are more representative than the classical student sample that is used in standard laboratory experiments. Indeed, although it is known that American MTurkers are still less representative than the general population (e.g., Asians are overrepresented and Blacks and Hispanics are underrepresented), it is also known that they are more representative than student samples (Berinsky et al., 2012; Paolacci and Chandler, 2014; Shapiro et al., 2013).

As every experimental work, also this one has some limitations. First of all, it explores the interaction between gender and cognitive reflection only in the case of black lies. It is possible that this effect changes when using lies with positive consequences for the receiver (altruistic white lies) or for both players (Pareto white lies), the rational being that people with high CRT might be more likely to be affected by the consequences of lying than people with low CRT, who might be more likely to follow heuristics and general rules of behavior that are independent of the specific context at hand (Toplak, West & Stanovich, 2011). Another limitation is that we look only at the cognitive reflection trait and not at the cognitive reflection state. For example, exploring whether gender and time pressure/delay interact in case of honesty is a fascinating direction for future research that is not covered by this work. Another potential limitation of this work is that we cannot be completely sure that our results are driven by gender differences in the interaction between lying aversion and cognitive reflection. Another possible explanation is that low cognitive reflection makes males but not females simply more likely to prefer the economic allocation ($0.50, $0.40) over the economic allocation ($0.40, $0.50). Although we cannot exclude this alternative interpretation for sure, we note that, as we have discussed in the Methods section, we have designed our experiment in such a way to minimize the effect of this potential confound: previous research found no evidence that low cognitive reflection makes men but not women more self-regarding or more spiteful (See Table 3 and Table 4 in the Supplementary Information of Capraro et al, 2017).

In sum, our results demonstrate that low cognitive reflection predicts honesty for men but not for women. This effect is driven by men that make choices in the Cognitive Reflection Test that are neither intuitive nor deliberative.

# References

Abeler, J., Nosenzo, D., & Raymond, C. (in press). Preferences for truth-telling. *Econometrica*.

Albaity, M., Rahman, M., & Shahidul, I. (2014). Cognitive reflection test and behavioral biases in Malaysia. *Judgment and Decision Making*, 9, 149-151.

Arechar, A. A, Gächter, S., & Molleman, L. (2018). Conducting interactive experiments online. *Experimental Economics*, 21, 99-131.

Barcelo, H., & Capraro, V. (2017). The Good, the Bad, and the Angry: An experimental study on the heterogeneity of people's (dis)honest behavior. *Available at SSRN: https://ssrn.com/abstract=3094305*.

Berinsky, A. J., Huber, G. A., & Lenz, G. S. (2012). Evaluating online labor markets for experimental research: Amazon.com's mechanical turk. *Political Analysis*, 20, 351-368.

Biziou-van-Pol, L., Haenen, J., Novaro, A., Occhipinti-Liberman, A., & Capraro, V. (2015). Does telling white lies signal pro-social preferences? *Judgment and Decision Making*, 10, 538-548.

Brañas-Garza, P., Capraro, V., & Rascón-Ramírez, E. (in press). Gender differences in altruism on Mechanical Turk: Expectations and actual behaviour. *Economics Letters*.

Brañas-Garza, P., Kujal, P., & Lenkei, B. (2015). Cognitive reflection test: Whom, how, when. *Available at https://mpra.ub.uni-meunchen.de/68409/1/MPRA_paper_68409.pdf*

Campitelli, G., & Gerrans, P. (2014). Does the cognitive reflection test measure cognitive reflection? A mathematical modeling approach. *Memory & Cognition*, 42, 434-447.

Cappelen, A. W., Sørensen, E. Ø., & Tungodden, B. (2013). When do we lie? *Journal of Economic Behavior and Organization*, 93, 258-265.

Capraro, V. (2018). Gender differences in lying in sender-receiver games: A meta-analysis. *Judgment and Decision Making*, 13, 345-355.

Capraro, V. (2017). Does the truth come naturally? Time pressure increases honesty in deception games. *Economics Letters*, 158, 54-57.

Capraro, V., & Cococcioni, G. (2015). Social setting, intuition, and experience in laboratory experiments interact to shape cooperative decision-making. *Proceedings of the Royal Society: Biological Sciences*, 282, 2015 0237.

Capraro, V., Corgnet, B., Espín, A. M., & Hernán-González, R. (2017). Deliberation favours social efficiency by making people disregard their relative shares: Evidence from USA and India. *Royal Society Open Science*, 4, 160605.

Capraro, V., Schulz, J., & Rand, D. G. (2018). Time pressure increases honesty in a sender-receiver deception game. *Available at SSRN: https://ssrn.com/abstract=3184537*.

Childs, J. (2012). Gender differences in lying. *Economics Letters*, 114, 147-149.

Croson, R. and Gneezy, U. (2009). Gender differences in preferences. *Journal of Economic Literature*, 47, 448–474.

Cueva, C., Iturbe-Ormaetxe, I., Mata-Pérez, E., Ponti, G., Sartarelli, M., Yu, H., & Zhukova, V. (2016). Cognitive (ir)reflection: New experimental evidence. *Journal of Behavioral and Experimental Economics*, 64, 81-93.

Dreber, A., & Johannesson, M. (2008). Gender differences in deception. *Economics Letters*, 99, 197-199.

Erat, S., & Gneezy, U. (2012). White lies. *Management Science*, 58, 723-733.

Federal Bureau of Investigation. (2018). Insurance Fraud. *Available at https://www.fbi.gov/stats-services/publications/insurance-fraud* [Accessed on Apr 3, 2018].

Fischbacher, U. Föllmi-Heusi, F. (2014). Lies in disguise – An experimental study on cheating. *Journal of the European Economic Association*, 11, 525-547.

Fosgaard, T., Hansen, L. G., Piovesan, M. (2013). Separating will from grace: An experiment on conformity and awareness in cheating. *Journal of Economic Behavior and Organization*, 93, 279-284.

Frederick, S. (2005). Cognitive reflection and decision making. *Journal of Economic Perspectives*, 194, 25-42.

Friesen, L., & Gangadharan, L. (2012). Individual level evidence of dishonesty and the gender effect. *Economics Letters*, 117, 624-626.

Gino, F., & Ariely, D. (2012). The dark side of creativity: Original thinkers can be more dishonest. *Journal of Personality and Social Psychology*, 102, 445-459.

Gino, F., Schweitzer, M. E., Mead, N. L., & Ariely, D. (2011). Unable to resist temptation: How self-control depletion promotes unethical behavior. *Organizational Behavior and Human Decision Processes*, 115, 191-203.

Goodman, J. K., Cryder, C. E., & Cheeman, A. (2013). Data collection in a flat world: The strengths and weaknesses of Mechanical Turk samples. *Journal of Behavioral Decision Making*, 26, 213-224.

Gneezy, U. (2005). Deception: The role of consequences. *The American Economic Review*, 95, 285-292.

Gneezy, U., Rockenbach, B., Serra-Garcia, M. (2013). Measuring lying aversion. *Journal of Economic Behavior and Organization*, 93, 293-300.

Gunia, B. C., Wang, L., Huang, L., Wang, J. W., Murnighan, J. K. (2012). Contemplation and conversation: subtle influences on moral decision making. *Academy of Management Journal*, 55, 13-33.

Herman, T. (2005). Study suggests tax cheating is on the rise. *The Wall Street Journal.* March 30.

Herman, T. (2005) Study suggests tax cheating is on the rise. *The Wall Street Journal*, March 30.

Horton, J. J., Rand, D. G., & Zeckhauser, R. J. (2011). The online laboratory: Conducting experiments in a real labor market. *Experimental Economics*, 14, 399-425.

Lohse, T., Simon, S. A., & Konrad, K. A. (2018). Deception under time pressure: Conscious decision or a problem of awareness? *Journal of Economic Behavior and Organization*, 146, 31-42.

Mazar, N., Amir, O., Ariely, D. (2008). The dishonesty of honest people: A theory of self-concept maintenance. *Journal of Marketing Research*, 45, 633-644.

Merkel, A. L., & Lohse, J. (2018). Is fairness intuitive? An experiment accounting for subjective utility differences under time pressure. *Experimental Economics*.

Paolacci, G., Chandler, J., & Ipeirotis, P. G. (2010). Running experiments on Amazon Mechanical Turk. *Judgment and Decision Making*, 5, 411-419.

Paolacci, G., & Chandler, J. (2014). Inside the Turk: Understanding Mechanical Turk as a participant pool. *Current Directions in Psychological Science*, 23, 184-188.

Pennycook, G., Cheyne, J. A., Koehler, D. J., & Fugelsang, J. A. (2016). Is the cognitive reflection test a measure of both reflection and intuition? *Behavior Research Methods*, 48, 341-348.

Primi, C., Morsanyi, K., Chiesi, F., Donati, M. A., & Hamilton, J. (2016). The development and testing of a new version of the cognitive reflection test applying item response theory (IRT). *Journal of Behavioral Decision Making*, 29, 453-469.

Rand, D. G. (2016). Social dilemma cooperation (unlike dictator game giving) is intuitive for men as well as women. *Journal of Experimental Social Psychology*, 73, 164-168.

Rand, D. G., Brescoll, V. L., Everett, J. A. C., Capraro, V., & Barcelo, H. (2016). Social heuristics and social roles: Intuition favors altruism for women but not for men. *Journal of Experimental Psychology: General*, 145, 389-396.

Rand, D. G., Greene, J. D., & Nowak, M. A. (2012). Spontaneous giving and calculated greed. *Nature*, 489, 427-430.

Ring, P., Neyse, L., David-Barett, T., & Schmidt, U. (2016). Gender differences in performance predictions: Evidence from the cognitive reflection test. *Frontiers in Psychology*, 7, 1680.

Ruffle, B. J., & Tobol, Y. (2017). Clever enough to tell the truth. *Experimental Economics*, 20, 130-155.

Shalvi, S., Eldar, O., & Bereby-Meyer, Y. (2012). Honesty requires time (and lack of justification). *Psychological Science*, 23, 1264-1270.

Shapiro, D. N., Chandler, J., & Mueller, P. A. (2013). Using Mechanical Turk to study clinical populations. *Clinical Psychological Science*, 1, 213–220.

Sinayev, A., & Peters, E. (2015). Cognitive reflection vs. calculation in decision making. *Frontiers in Psychology*, 6.

Sutter, M. (2009). Deception through telling the truth? Experimental evidence from individuals and teams. *Economic Journal*, 119, 47-60.

Toplak, M. E., West, R. F., & Stanovich, K. E. (2011). The Cognitive Reflection Test as a predictor of performance in heuristics-and-biases tasks. *Memory & Cognition*, 39, 1275.

Toplak, M. E., West, R. F., & Stanovich, K. E. (2014). Assessing miserly information processing: An expansion of the Cognitive Reflection Test. *Thinking & Reasoning*, 20, 147-168.

van't Veer, A. E., Stel, M., van Beest, I. (2014). Limited capacity to lie: Cognitive load interferes with being dishonest. *Judgment and Decision Making*, 9, 199-206.

Verschuere, B., Köbis, N. C., Bereby-Meyer, Y., Rand, D. G., & Shalvi, S. (2018). Taxing the brain to uncover lying? Meta-analyzing the effect of imposing cognitive load on the reaction-time cost of lying. *Journal of Applied Memory & Cognition*.

Verschuere, B., & Shalvi, S. (2014). The truth comes naturally! Does it? *Journal of Language and Social Psychology*, 33, 417-423.

**Appendix**

*Experimental instructions of the deception game*

Please read these instructions carefully. You may earn a considerable sum of money, depending on the decisions you make.

You have been matched with another participant. Your bonus and that of the other participant depends only on your choice. The other participant does not play any active role.

There are two groups (group 1, group 2). You will be randomly assigned to one of them. You will be informed of the group you have been assigned to, but the other participant will not.

We will ask you to declare the group you have been assigned to. So you get to choose between two possible options:

Option 1: "I have been assigned to group 1."
Option 2: "I have been assigned to group 2."

If you choose the option corresponding to the number of the actual group you have been assigned to, then you will receive 40c and the other participant will receive 50c.

If you choose the option corresponding to the number of the other group (the one you have not been assigned to), then you will receive 50c and the other participant will receive 40c.

Finally, only you will be informed of the particular monetary value connected to each message. The other participant will not be informed of these monetary values.

Here are some questions to ascertain that you understand the rules. Remember that you have to answer all of these questions correctly in order to get the completion code. <u>If you fail any of them, the survey will automatically end and you will not get any payment.</u>

What is the choice that maximises YOUR outcome?
(Available choices: Choosing the message corresponding to the number of the actual group you have been assigned to/ Choosing the message corresponding to the number of the other group (the one you have not been assigned to))

What is the choice that maximises the OTHER PARTICIPANT'S outcome?
(Available choices: Choosing the message corresponding to the number of the actual group you have been assigned to/ Choosing the message corresponding to the number of the other group (the one you have not been assigned to))

Congratulations, you have passed all comprehension questions. It is now time to make your choice.

You have been assigned to group 1 (group 2, random allocation).

Which option do you choose?

(Available choices: "I have been assigned to group 1"/"I have been assigned to group 2")

*Experimental instructions of the CRT*

Please answer the following questions

A jar of Peanut Butter and a jar of Jam cost $10.20 in total. The jar of Peanut Butter costs $10.00 more than the jar of Jam. How much does the jar of Jam cost? _____ cents
(choice to be typed in a blank text box)

If it takes 6 machines 6 minutes to make 6 widgets, how long would it take 120 machines to make 120 widgets? _____ minutes
(choice to be typed in a blank text box)

In a lake, there is a patch of lily pads. Every day, the patch doubles in size. If it takes 50 days for the patch to cover the entire lake, how long would it take for the patch to cover half of the lake? _____ days
(choice to be typed in a blank text box)